\documentclass[dvipsnames,format=sigconf]{acmart}

\usepackage{amssymb}
\usepackage{amsmath}
\usepackage{relsize}
\usepackage{xcolor}
\usepackage{url}
\usepackage{graphicx}
\usepackage{multirow}
\usepackage{multicol}
\usepackage{comment}
\usepackage{booktabs} 
\usepackage{array} 
\usepackage{balance} 
\usepackage{soul}
\usepackage{makecell}
\usepackage[skip=1pt]{caption}

\AtBeginDocument{%
  }

\copyrightyear{2025} 
\acmYear{2025} 
\setcopyright{cc}
\setcctype{by}
\acmConference[GECCO '25 Companion]{Genetic and Evolutionary Computation Conference}{July 14--18, 2025}{Malaga, Spain}
\acmBooktitle{Genetic and Evolutionary Computation Conference (GECCO '25 Companion), July 14--18, 2025, Malaga, Spain}
\acmDOI{10.1145/3712255.3734313}
\acmISBN{979-8-4007-1464-1/2025/07}

\begin{document}

\title{Steiner Traveling Salesman Problem with Quantum Annealing}
\author{Alessia Ciacco}
\affiliation{%
  \institution{Department of Mechanical, Energy and Management Engineering, University of Calabria}
  \city{Rende}
  \country{Italy}
}
\email{alessia.ciacco@unical.it}

\author{Francesca Guerriero}
\affiliation{%
  \institution{Department of Mechanical, Energy and Management Engineering, University of Calabria}
  \city{Rende}
  \country{Italy}
}
\email{francesca.guerriero@unical.it}

\author{Eneko Osaba}
\affiliation{%
  \institution{TECNALIA, Basque Research and Technology Alliance (BRTA)}
  \streetaddress{Astondo Bidea, Edificio 700}
  \city{Derio}
   \country{Spain}
}
\email{eneko.osaba@tecnalia.com}

\renewcommand{\shortauthors}{Ciacco et al.}

\begin{abstract}
The Steiner Traveling Salesman Problem (STSP) is a variant of the classical Traveling Salesman Problem. The STSP involves incorporating \textit{steiner} nodes, which are extra nodes not originally part of the required visit set but that can be added to the route to enhance the overall solution and minimize the total travel cost. Given the NP-hard nature of the STSP, we propose a quantum approach to address it. Specifically, we employ quantum annealing using D-Wave's hardware to explore its potential for solving this problem. To enhance computational feasibility, we develop a preprocessing method that effectively reduces the network size. Our experimental results demonstrate that this reduction technique significantly decreases the problem complexity, making the Quadratic Unconstrained Binary Optimization formulation, the standard input for quantum annealers, better suited for existing quantum hardware. Furthermore, the results highlight the potential of quantum annealing as a promising and innovative approach for solving the STSP.
\end{abstract}



\maketitle

\section{Introduction}
The Traveling Salesman Problem (TSP, \cite{dantzig1954solution}) is one of the most extensively studied combinatorial optimization problems, with numerous applications in fields such as logistics, transportation planning, and industrial manufacturing. The classical formulation of the TSP requires determining the shortest route for a salesman to visit a set of given locations and return to the starting point. Specifically, as the number of cities increases, the number of possible routes grows exponentially, making exact solutions impractical for large instances. The TSP is classified as an NP-hard problem, which means that no known algorithm can solve it in polynomial time for all instances. This complexity has driven the development of a wide range of solution approaches with the aim of finding optimal or near-optimal solutions within a reasonable time frame.

Similarly, the Steiner Tree Problem (STP, {\cite{gilbert1968steiner}) is a combinatorial optimization problem that involves finding a minimum-cost tree that connects a specific subset of nodes (called \textit{terminal} nodes) in a graph, while also being able to add other nodes (\textit{steiner} nodes) to reduce the overall cost. The STP has applications in telecommunications networks, circuit design, computational biology, and logistics. Because the STP is NP-hard, approximations and heuristics are used to solve it in practical cases.

Building on these concepts, the Steiner Traveling Salesman Problem (STSP) represents a variant combining both the TSP and the STP. In this problem, a traveling salesperson must find the minimum cost path that visits all terminal nodes of a graph at least once, but can also use steiner nodes (not mandatory to visit) to reduce the overall path cost. The STSP allows the inclusion of additional nodes and permits edges to be crossed more than once, enhancing path efficiency.
This feature makes the STSP particularly relevant for real-world applications, such as the design of communication networks or transportation systems, where utilizing existing infrastructure through steiner nodes can lead to significantly more cost-effective and practical solutions.

The STSP was first introduced by Cornuéjols et al. in \cite{cornuejols1985traveling} and Fleischmann in \cite{fleischmann1985cutting}. Later, Letchford et al. proposed in  \cite{letchford2013compact} a mathematical formulations for the STSP by adapting existing models for the TSP. They introduced two single-commodity flow models, a multi-commodity flow model, and two temporal formulations for the STSP on directed graphs. More recently, Rodríguez-Pereira et al. presented in \cite{rodriguez2019steiner} a compact integer linear programming (ILP) formulation, leveraging the property that an optimal solution to the STSP exists when no edge is traversed more than twice. Like the TSP and STP, the STSP is classified as NP-hard, making exact solutions computationally prohibitive for large instances.

In a different vein, quantum computing has emerged as an innovative and promising approach to solving combinatorial optimization problems, particularly those classified as NP-hard. Thanks to quantum mechanical principles such as superposition and entanglement, quantum algorithms can explore many possible solutions simultaneously, increasing the potential to find optimal configurations in less time than traditional methods. For a comprehensive survey on quantum approaches used to solve routing problems with quantum algorithms, the reader is referred to \cite{ciacco2025review} and \cite{osaba2022systematic}.

Among the various paradigms of quantum algorithms, Quantum Annealing (QA) has attracted special interest as one of the most promising methods for efficiently solving complex combinatorial optimization problems. QA takes its name from an analogous physical process, annealing, which involves a material being heated and then cooling down slowly. This process allows the material to settle back into a state of low energy, ultimately reaching a low-energy location in a solution space. Additionally, quantum mechanical properties are used to explore the possible states of the material.

In practical applications, many optimization problems can be formulated as Quadratic Unconstrained Binary Optimization (QUBO) problems \cite{glover2018tutorial}, which can be naturally mapped onto the hardware of quantum annealers, allowing them to efficiently search for near-optimal solutions. One prominent example of such quantum devices is the ones built by D-Wave Systems, which enable QA at a scale that has been used to solve many real-world optimization problems.

This study is the first to apply a quantum computing approach to solving the STSP. Specifically, we introduce the use of QA for this task, representing a pioneering effort in applying quantum methods to such a complex optimization problem, such as STSP}. In particular, we analyze QA by using two different approaches: the D-Wave Quantum Processing Unit (QPU) and the D-Wave's \texttt{LeapBQM} hybrid solver \cite{osaba2024hybrid}. We provide a comprehensive evaluation of these two distinct quantum approaches to tackle the STSP. In addition to this, we introduce a network reduction method to enhance the QUBO representation of the problem, reducing the number of variables that would otherwise grow significantly when transforming the ILP formulation into QUBO. This formulation constitutes another significant contribution of this research.

The remaining of this paper is structured as follows. Section~\ref{sec:problem_statement} presents the problem description, the mathematical formulation and the proposed preprocessing method for reducing arcs.  Section~\ref{sec:resolutive approaches} describes the proposed solution approach for the problem.  Section~\ref{sec:experiments} presents the computational tests and analysis of the results. Section~\ref{sec:conclusions} summarizes the concluding remarks.

\section{Problem statement}\label{sec:problem_statement}
This section presents the characteristics of the STSP addressed in this work and is divided into three subsections. The subsection \ref{subsec:characteristics_of_problem} defines the problem, detailing the sets and parameters involved. The subsection \ref{subsec:Mathematical Formulation} defines the variables, presents the mathematical formulation of the problem, and explains the objective function and constraints. Finally, the subsection \ref{sezione_metodo_archi} outlines the preprocessing method used to reduce the number of arcs.

\subsection{Characteristics of the problem}\label{subsec:characteristics_of_problem}
\begin{figure}[htbp] 
\centering
    \includegraphics[width=0.48\textwidth]{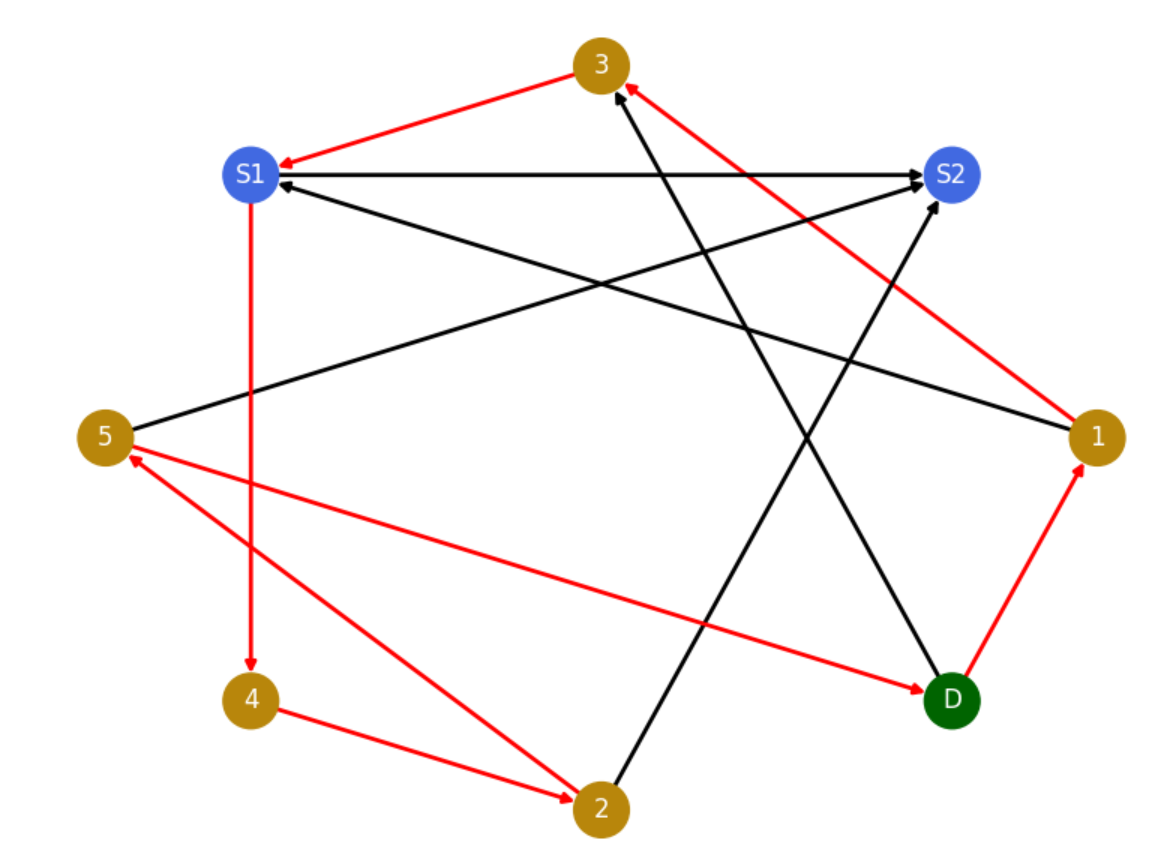}
    \caption{Representation of the STSP network: the yellow circles represent the $V_R$ nodes, the green circle indicates the depot and the blue circles represent the steiner nodes. The black edges represent the arcs of the graph, while a possible route is highlighted in red, showing one of the potential solutions to the optimization problem.}
    \Description{Graph representation of the STSP network. The graph consists of yellow, blue, and green circles connected by black and red directed edges. The yellow circles represent the $V_R$ nodes, the green circle is the depot, and the blue circles are the Steiner nodes. The black edges represent the arcs of the graph, while a possible route is highlighted in red, illustrating one potential solution to the optimization problem.}
    \label{stsp}
\end{figure}

The STSP is defined on a directed graph $G = (V, A)$, where $V$ is the set of vertices and $A$ is the set of directed arcs. Let $V_R \subseteq V$ be the terminal nodes, which must be visited. The remaining nodes, $V \setminus V_R$, are called steiner, which are vertices that can be strategically included to optimize the route. These steiner nodes can help minimize the total travel distance by creating more efficient connections between terminal nodes. Figure~\ref{stsp} illustrates the representation of this network. The yellow points represent the $V_R$ vertices, the green dot represents the depot and the blue vertices represent the steiner nodes. 

Let $\delta^+(i)$ the set of arcs outgoing from node $i$, i.e., the set of directed edges that originate from node $i$ and point to other nodes in the graph. Let $\delta^-(i)$ be the set of arcs incoming to node $i$, i.e., the set of directed edges that point to $i$ from other nodes in the graph.

Each directed arc $k \in A$ is associated with a specific cost $c_k$, which represents the cost of traversing that arc. The goal is to find a route that minimizes the total traversal cost while ensuring that every node in the terminal set $V_R$ is visited at least once. 
The parameters used in the formulation of the STSP are summarized in Table~\ref{tab:notations}.
\begin{table}[h]
\centering
\caption{Summary of the notations used in the model formulation.}
\begin{tabular}{c|p{6cm}}
\hline
\textbf{Notation} & \textbf{Description} \\ \hline
$V$               & Set of all vertices in the graph \\
$A$               & Set of directed arcs in the graph \\
$V_R$             & Set of the required nodes to visit \\
$\delta^+(i)$     & Set of arcs outgoing from node $i$ \\
$\delta^-(i)$     & Set of arcs incoming to node $i$ \\
$c_k$             & Cost associated with traversing arc $k$ \\
\hline
\end{tabular}
\label{tab:notations}
\end{table}

\subsection{Mathematical Formulation}\label{subsec:Mathematical Formulation}
The variable used to define the model is:

\begin{description}
    \item[$y_k^t$] $\begin{cases} 
    1, & \text{if arc } k \text{ is traversed at period } t \\
    0, & \text{otherwise}
    \end{cases}$
\end{description}

The formulation presented in this paper is inspired by the model proposed by Letchford et al. in \cite{letchford2013compact}.  
\begin{align}
    &\text{min} \quad \sum_{t=1}^{|A|} \sum_{k \in A} c_k \;y_k^t, \label{eq:objective} \\
    &\text{s.t.} \nonumber \\
    &\sum_{k \in \delta^+(0)} y_k^1 = 1 \quad  \label{eq:depot_start}\\
    & y_k^{1} = 0, \quad \forall k \in A \setminus \delta^+(0) \label{eq:depot_end} \\
    &\sum_{t=1}^{|A|} \sum_{k \in \delta^+(0)} y_k^t = \sum_{t=1}^{|A|} \sum_{k \in \delta^-(0)} y_k^t \label{eq:visit_once} \\
    & \sum_{t=1}^{|A|} \sum_{k \in \delta^+(i)} y_k^t \geq 1, \quad \forall i \in V_R \label{eq:visit_required}\\
    &\sum_{k \in \delta^-(i)} y_k^t = \sum_{k \in \delta^+(i)} y_k^{t+1}, \quad \forall i \in V\setminus\{0\}, 1 \leq t \leq |A|-1 \label{eq:flow_conservation} \\
    &y_k^t \in \{0, 1\}, \quad \forall k \in A, 1 \leq t \leq |A| \label{eq:binary}
\end{align}
The objective function (\ref{eq:objective}) minimizes the total travel cost across all time periods. It is expressed as the sum of the costs associated with each arc traversed, weighted by the respective cost.
Constraint (\ref{eq:depot_start}) ensures that the route starts at the depot. 
Constraints (\ref{eq:depot_end}) ensure that in the first period, there are no other arcs traversed except for the one originating from the depot.
Constraint (\ref{eq:visit_once}) ensures that the sum of incoming arcs at the depot node equals the sum of outgoing arcs.
Constraints (\ref{eq:visit_required}) enforce the requirement that all required vertices  must be visited at least once. 
Constraints (\ref{eq:flow_conservation}) ensure flow conservation between consecutive time periods for each node (except the depot). They state that the number of incoming arcs to a vertex in any given period must equal the number of outgoing arcs from that node in the next period, thereby maintaining a continuous flow through the network. Constraints (\ref{eq:binary}) define the domain of the decision variable. It ensures that the index $k$ belongs to the set $A$ of arcs and that the time index $t$ is defined in the interval $1 \leq t \leq |A|$. The upper bound $|A|$ guarantees that the model can represent even the longest path, namely the one that traverses all arcs in the graph.

\subsection{Preprocessing method for reducing arcs} \label{sezione_metodo_archi}
In order to enhance the efficiency of problem formulation and reduce its computational complexity, we have developed a preprocessing method to reduce the size of the set $A$. 
The \textit{Preprocessing Method for Reducing Arcs} (PMRA) reduces the size of the arc set by removing unnecessary or excessively costly connections, simplifying the optimization task and improving computational efficiency. 

This reduction is particularly important when considering the context of quantum computing. Currently, we are immersed the \textit{noisy intermediate-scale quantum} (NISQ, \cite{preskill2018quantum}) era, a period characterized by quantum computers' limitations in efficiently handling problems, even those of small to medium size. For this reason, by reducing the size of the problem, we can tackle larger-scale optimization problems that would otherwise be too complex or infeasible for current quantum devices. Furthermore, reducing the problem size enables quantum algorithms to prioritize core variables and constraints, enhancing the efficiency for smaller problems. Thus, by simplifying the model, we can not only improve the performance in solving smaller instances with greater precision but also enable the resolution of larger instances that would otherwise be unsolvable without this approach.


The preprocessing method follows these steps:
\begin{enumerate}
    \item \textbf{Eliminate irrelevant arcs:} remove all arcs that do not have at least one terminal node from the set $\{V_R\cup0\}$
        \item \textbf{Cost calculation and threshold definition:} compute the average cost $m$ from the reduced set of arcs. Define a threshold value $\alpha$ as a 10\% increase over the average cost, i.e., $\alpha = m + 0.1m$.
            \item \textbf{Removal of high-cost arcs:} remove all arcs whose cost is greater than or equal to the threshold value $\alpha$, provided that none of the following conditions hold:
    \begin{itemize}
        \item The arc $(i, j)$ belongs to the set $\{V_R\cup0\}$;
           \item The removal of the arc does not isolate any node in the graph, i.e., no vertices becomes disconnected from all other nodes.
      
    \end{itemize}
    \item \textbf{Elimination of isolated steiner nodes:} identify steiner nodes that have become either completely disconnected (i.e., with no incoming and outgoing arcs) or weakly connected (i.e., with no incoming or outgoing arcs). These nodes, along with any associated arcs, are removed. This process is applied recursively until no steiner nodes remain in an isolated or weakly connected state.
\end{enumerate}


\section{Solution approach} \label{sec:resolutive approaches}
We adopt a structured, multi-phase approach to solve the STSP using QA. The process begins by formulating the problem as an ILP. To make the problem more tractable we introduce the PMRA. To assess its performance, we resolve the ILP problem with and without PMRA using Gurobi. Subsequently, the problem is converted into QUBO formulation. The QUBO formulation, both with and without PMRA, is solved using the Simulated Annealing (SA) algorithm and the QA algorithm.

\begin{table*}[ht]
\caption{Results of the STSP problem, solved with Gurobi, with and without PMRA. It presents the average and standard deviation of the objective function (total cost) or execution time and the percentage of solved instances. The symbol ``-" indicates that no solution is found in any of the 10 executions.} \label{QUBO}
\centering
\resizebox{0.85\textwidth}{!}{%
\begin{tabular}{c|cc>{\centering\arraybackslash}p{1.5cm}|cc|cc>{\centering\arraybackslash}p{1.5cm}|ccc}
\textbf{$|V|$} & \multicolumn{3}{c|}{\makecell{\textbf{SILP }\\ \textbf{Objective Function}}} & \multicolumn{2}{c|}{\makecell{\textbf{SILP }\\ \textbf{Time}}} & \multicolumn{3}{c|}{\makecell{\textbf{RILP }\\ \textbf{Objective Function}}} & \multicolumn{2}{c}{\makecell{\textbf{RILP }\\ \textbf{Time}}} \\ \hline
 & Avg & Stan.Dev. & \%Sol.Inst. & Avg & Stan.Dev. & Avg & Stan.Dev. & \%Sol.Inst. & Avg & Stan.Dev. \\ \hline

 4     & \textbf{122,03} & 0,00  & 100\% & \textbf{0,00} & 0,00  & \textbf{122,03} & 0,00  & 100\% & \textbf{0,00}  & 0,00 \\
    5     & \textbf{90,14} & 0,00  & 100\% & 0,01 & 0,00  & \textbf{90,14} & 0,00  & 100\% & \textbf{0,00}  & 0,00 \\
    6     & \textbf{155,00} & 0,00  & 100\% & 0,02 & 0,00  & \textbf{155,00} & 0,00  & 100\% & \textbf{0,01}  & 0,00 \\
    7     & \textbf{157,52} & 0,00  & 100\% & 0,05 & 0,01  & \textbf{157,52} & 0,00  & 100\% & \textbf{0,03}  & 0,01 \\
    8     & \textbf{147,79} & 0,00  & 100\% & 0,03 & 0,01  & \textbf{147,79} & 0,00  & 100\% & \textbf{0,01}  & 0,00 \\
    9     & \textbf{179,84} & 0,00  & 100\% & 0,53 & 0,11  & \textbf{179,84} & 0,00  & 100\% & \textbf{0,10}  & 0,02 \\
    10    & \textbf{181,42} & 0,00  & 100\% & 0,11 & 0,04  & \textbf{181,42} & 0,00  & 100\% & \textbf{0,05}  & 0,01 \\
    11    & \textbf{218,70} & 0,00  & 100\% & 2,51 & 0,37  & \textbf{218,70} & 0,00  & 100\% & \textbf{1,71}  & 0,26 \\
    12    & \textbf{210,53} & 0,00  & 100\% & 2,03 & 0,41  & \textbf{210,53} & 0,00  & 100\% & \textbf{1,68}  & 0,17 \\
    13    & \textbf{244,77} & 0,00  & 100\% & 7,91 & 3,64  & \textbf{244,77} & 0,00  & 100\% & \textbf{4,24}  & 0,38 \\
    14    & \textbf{253,74} & 0,00  & 100\% & 1,14 & 0,28  & \textbf{253,74} & 0,00  & 100\% & \textbf{0,62}  & 0,18 \\
    15    & \textbf{257,72} & 0,00  & 100\% & 1,19 & 0,29  & \textbf{257,72} & 0,00  & 100\% & \textbf{0,56}  & 0,19 \\
    16    & -     & -     & 0\%   & 8,20 & 3,80  & \textbf{308,70} & 0,00  & 100\% & \textbf{6,03}  & 0,41 \\
    17    & -     & -     & 0\%   & 9,38 & 0,23  & \textbf{295,36} & 0,00  & 100\% & \textbf{1,58}  & 0,37 \\
    18    & -     & -     & 0\%   & 9,60 & 0,30  & \textbf{268,42} & 0,00 & 90\%  & \textbf{6,79}  & 1,91 \\
    19    & -     & -     & 0\% &  9,46 & 0,28  & -     & -     & 0\%   & 9,50  & 0,29 \\

\end{tabular}}

\end{table*}
The solution process is structured as follows:
\begin{enumerate}
    \item \textbf{Mathematical Formulation}: we formulate the problem using the ILP model;
    
    \item \textbf{PMRA}: we apply the PMRA to reduce the complexity of the problem;
    \item \textbf{ILP scalability analysis}: we solve the ILP formulation with and without PMRA using Gurobi to evaluate the scalability of this method. The ILP solved without PMRA, \textit{Standard ILP} (SILP), serves as a benchmark, while the ILP solved with PMRA, \textit{Reduced ILP (RILP)}, applies the proposed reduction approach;
    \item \textbf{QUBO formulation}: we convert both the preprocessed and unpreprocessed versions of the STSP into QUBO formulations. The preprocessed version will be referred to as \textit{Reduced QUBO (RQUBO)}, while the unprocessed version will be referred to as \textit{Standard QUBO (SQUBO)}. The QUBO formulation is solved using two different approaches:
    \begin{itemize}
        \item \textbf{SA solution}: we solve both SQUBO and RQUBO problems using the SA algorithm. 
        \item \textbf{QA solution}: 
    we apply QA to solve both the SQUBO and RQUBO problems. Specifically, we use both the QPU-based approach and the \texttt{LeapBQM} hybrid approach. The QPU approach directly utilizes the quantum properties of the system, while the hybrid approach combines quantum and classical resources to enhance performance, allowing for a broader range of problem sizes and more efficient solutions.
\end{itemize}
\end{enumerate}

\section{Computational study} \label{sec:experiments} 
All computations are performed on an Intel Xeon 2.20GHz with 12GB of RAM. We have conducted 10 tests for each instance in order to evaluate the performance and robustness of the model.

For the instances solved using Gurobi, we have used the Gurobi \textit{ver. 11} \cite{gurobi-doc} and the implementation is done entirely in Python. Also, we impose a \texttt{time\_limit} of 10 seconds. For SILP and RILP solved with Gurobi, we consider 16 instances with increasing sizes, ranging from 4 to 19 nodes. The analysis stops beyond this threshold, as no solution is obtained within the time limit. 

Furthermore, we have used the \texttt{dimod.cqm\_to\_bqm} method from the \texttt{Ocean SDK} library to convert the ILP model, which includes constraints, into a Binary Quadratic Model. This conversion transforms the constraints into penalty terms, resulting in an unconstrained model that can be further processed as a QUBO formulation for quantum optimization. It is important to note that this conversion alters the structure of the problem, and due to the introduction of bias and penalty terms, the solutions obtained from the ILP and QUBO formulations will have different energy values.

For the SA, we employ the \texttt{neal.sampler.SimulatedAnnealing Sampler} implementation of the D-Wave's Neal library. The SA sampler is executed with the \texttt{num\_reads} parameter set to 1000. The remaining parameters are kept at their default values, following Neal’s documentation. Specifically, the \texttt{beta\_range} parameter, which defines the temperature range for the solver, is automatically calculated based on the total bias associated with each problem node. During the annealing process, it is interpolated according to the mode specified by the \texttt{beta\_schedule\_type} parameter, and the default mode uses geometric interpolation. For SQUBO and RQUBO solved with SA, we consider 4 instances, ranging from 4 to 7 nodes. The analysis stops beyond this threshold, as no solution is found for any of the 10 tests, neither for SQUBO nor RQUBO, within the time limit.

Finally, as previously mentioned, we have employed two distinct approaches for QA utilizing resources provided by D-Wave. The first approach is the D-Wave's \texttt{Advantage\_System7.1} QPU. The second is the \texttt{LeapBQM} hybrid solver, which combines the D-Wave QPU with classical algorithms to enhance overall efficiency, particularly for more complex problems that may not be efficiently solvable using the QPU alone. We refer readers interested on further details of the \texttt{LeapBQM} to \cite{HSS}.

Experimental tests are conducted within a 100x100 unit square region. The set of nodes $V$ is randomly generated by assigning each node random Cartesian coordinates within this area. An additional node, denoted as the depot node $0$, is generated with random coordinates within the same region. The subset of required nodes $V_R \subseteq V$ is selected comprising the first $70\%$ of the total vertices in $V$, rounded down to the nearest integer. Furthermore, each arc cost $c_k$ is assigned a random value  between 20 and 50. To facilitate the replication of this study, all instances used are publicly accessible in \cite{STSPData}.
\begin{table}[b]
    \caption{Comparison of the number of variables in SQUBO and RQUBO formulations.}
    \centering
    \begin{tabular}{>{\centering\arraybackslash}p{1.6cm} |>{\centering\arraybackslash}p{1.6cm}| >{\centering\arraybackslash}p{1.6cm}| >{\centering\arraybackslash}p{1.6cm}}
        \textbf{$|V|$} & \textbf{n.var SQUBO} & \textbf{n.var RQUBO} & \textbf{GAP} \\
        \hline
    4     & 156   & \textbf{91}    & 42\% \\
    5     & 414   & \textbf{208}   & 50\% \\
    6     & 697   & \textbf{308}   & 56\% \\
    7     & 1401  & \textbf{706}   & 50\% \\
    8     & 2637  & \textbf{1121}  & 57\% \\
    9     & 4401  & \textbf{2254}  & 49\% \\
    10    & 6949  & \textbf{4023}  & 42\% \\
    11    & 8906  & \textbf{4689}  & 47\% \\
    12    & 13533 & \textbf{6793}  & 50\% \\
    13    & 19132 & \textbf{9684}  & 49\% \\
    14    & 26343 & \textbf{14498} & 45\% \\
    15    & 36208 & \textbf{18868} & 48\% \\
    16    & 42969 & \textbf{22911} & 47\% \\
    17    & 55828 & \textbf{29710} & 47\% \\
    18    & 73032 & \textbf{35851} & 51\% \\
    19    & 91965 & \textbf{52585} & 43\% \\
    \end{tabular}
    \label{scalabilità}
\end{table}

\begin{table*}[ht]
\centering
\caption{Results of STSP problem for SA with SQUBO and RQUBO. It presents the average and standard deviation of the objective function (total cost) or execution time and the percentage of solved instances. The symbol ``-" indicates that no solution is found in any of the 10 executions.} \label{SA}
\resizebox{0.8\textwidth}{!}{%
\begin{tabular}{c|ccc|cc|ccc|cc}
\textbf{$|V|$} & \multicolumn{3}{c|}{\textbf{SA for SQUBO}} & \multicolumn{2}{c|}{\textbf{Time (SQUBO)}} & \multicolumn{3}{c|}{\textbf{SA for RQUBO}} & \multicolumn{2}{c}{\textbf{Time (RQUBO)}} \\ \hline
 & Avg & Std.Dev. & \%Sol.Inst. & Avg & Std.Dev. & Avg & Std.Dev. & \%Sol.Inst. & Avg & Std.Dev. \\ \hline
4 & 2151,13 & 380,89 & 100\% & 9,67 & 3,53 & \textbf{2065,22} & 2631,68 & 100\% & \textbf{4,57} & 0,39 \\
5 & 5870,31 & 428,74 & 30\% & 46,51 & 0,68 & \textbf{2837,97} & 291,93 & 100\% & \textbf{18,90} & 1,66 \\
6 & - & - & 0\% & 89,76 & 0,83 & - & - & 0\% & 26,55 & 0,29 \\
7 & - & - & 0\% & 320,69 & 8,00 & - & - & 0\% & 120,42 & 1,62 \\
\end{tabular}}
\end{table*}

We analyze the behavior of the considered problems, focusing on different aspects of resolution. The analysis of the numerical results is structured into three sections: 
\begin{itemize}  
    \item The first section examines the scalability of our proposed method;  
    \item The second section analyzes the outcomes of solving these QUBO problems using the SA algorithm;
    \item The third section explores the performance of QA by employing the D-Wave solver.  
\end{itemize}  
\begin{table*}[ht]
\caption{Results of the STSP problem for QA on a QPU, comparing SQUBO and RQUBO. It reports the mean and standard deviation of both the objective function (total cost) and execution time, along with the percentage of successfully solved instances. The symbol ``-'' indicates cases where no valid solution was found in any of the five runs.} 
\label{QA with QPU}
\centering
\resizebox{\textwidth}{!}{%
\begin{tabular}{>{\centering\arraybackslash}m{1.2cm}|>{\centering\arraybackslash}m{1.5cm}>{\centering\arraybackslash}m{1.5cm}>{\centering\arraybackslash}m{2cm}|%
>{\centering\arraybackslash}m{1.5cm}>{\centering\arraybackslash}m{1.5cm}|%
>{\centering\arraybackslash}m{1.5cm}>{\centering\arraybackslash}m{1.5cm}>{\centering\arraybackslash}m{2cm}|%
>{\centering\arraybackslash}m{1.5cm}>{\centering\arraybackslash}m{1.5cm}}
\textbf{$|V|$} & \multicolumn{3}{c|}{\makecell{\textbf{QA with QPU for SQUBO}\\ \textbf{Objective Function}}} & 
\multicolumn{2}{c|}{\makecell{\textbf{QA with QPU for SQUBO}\\ \textbf{Time}}} & 
\multicolumn{3}{c|}{\makecell{\textbf{QA with QPU for RQUBO}\\ \textbf{Objective Function}}} & 
\multicolumn{2}{c}{\makecell{\textbf{QA with QPU for RQUBO}\\ \textbf{Time}}} \\ \hline
 & Avg & Stan.Dev. & \%Sol.Inst. & Avg & Stan.Dev. & Avg & Stan.Dev. & \%Sol.Inst. & Avg & Stan.Dev. \\ \hline
    4     & 2864,14 & 384,93 & 100\% & 221,27 & 118,55 & \textbf{1591,52} & 147,36 & 100\% & \textbf{50,04} & 21,55 \\
    5     & -     & -     & 0\%   & -     & -     & \textbf{3484,61} & 258,19 & 100\% & \textbf{273,29} & 113,16 \\
    6     & -     & -     & 0\%   & -     & -     & - & - & - & - & - \\

\end{tabular}}
\end{table*}

\begin{table*}[ht]
\caption{Results of the STSP problem solved using the \texttt{LeapBQM} hybrid solver, with a comparison between SQUBO and RQUBO formulations. It includes the average and standard deviation of both the objective function (total cost) and execution time, as well as the percentage of instances successfully solved. A value of ``-'' indicates that no valid solution was obtained in any of the five runs.} \label{QA with hybrid solver}
\centering
\resizebox{\textwidth}{!}{%
\begin{tabular}{>{\centering\arraybackslash}m{1.2cm}|>{\centering\arraybackslash}m{1.5cm}>{\centering\arraybackslash}m{1.5cm}>{\centering\arraybackslash}m{2cm}|%
>{\centering\arraybackslash}m{1.5cm}>{\centering\arraybackslash}m{1.5cm}|%
>{\centering\arraybackslash}m{1.5cm}>{\centering\arraybackslash}m{1.5cm}>{\centering\arraybackslash}m{2cm}|%
>{\centering\arraybackslash}m{1.5cm}>{\centering\arraybackslash}m{1.5cm}}
\textbf{$|V|$} & \multicolumn{3}{c|}{\makecell{\textbf{QA with LeapBQM for SQUBO}\\ \textbf{Objective Function}}} & 
\multicolumn{2}{c|}{\makecell{\textbf{QA with LeapBQM for SQUBO}\\ \textbf{Time}}} & 
\multicolumn{3}{c|}{\makecell{\textbf{QA with LeapBQM for RQUBO}\\ \textbf{Objective Function}}} & 
\multicolumn{2}{c}{\makecell{\textbf{QA with LeapBQM for RQUBO}\\ \textbf{Time}}} \\ \hline
 & Avg & Stan.Dev. & \%Sol.Inst. & Avg & Stan.Dev. & Avg & Stan.Dev. & \%Sol.Inst. & Avg & Stan.Dev. \\ \hline
    4     & 1508,31 & 45,69 & 100\% & 2,42  & 0,97  & \textbf{529,95} & 64,02 & 100\% & \textbf{2,17}  & 0,92 \\
    5     & 3646,18 & 270,49 & 100\% & 2,52  & 0,94  & \textbf{2140,54} & 227,49 & 100\% & \textbf{2,26}  & 0,93 \\
    6     & 7589,27 & 773,18 & 100\% & 2,57  & 1,13  & \textbf{2685,72} & 998,30 & 100\% & \textbf{2,43}  & 1,05 \\
    7     & 15901,22 & 1382,77 & 100\% & 3,78  & 0,05  & \textbf{7791,49} & 616,05 & 100\% & \textbf{3,47}  & 0,11 \\
    8     & 29374,55 & 856,38 & 100\% & 1,13  & 0,08  & \textbf{12740,63} & 832,71 & 100\% & \textbf{0,87}  & 0,00 \\
    9     & 57309,63 & 2165,80 & 100\% & 2,32  & 0,15  & \textbf{28061,95} & 2907,76 & 100\% & \textbf{2,12}  & 0,09 \\
   
\end{tabular}}
\end{table*}
\subsection{Scalability of PMRA using Gurobi solver}
We analyze the scalability of our proposed method, focusing on how it performs as the problem size increases.
Table~\ref{QUBO} shows the results of solving the STSP problem with Gurobi, with and without PMRA. 
The table~\ref{QUBO} is divided into four part: \textit{SILP Objective Function}, \textit{SILP Time}, \textit{RILP Objective Function} and \textit{RILP Time}. Each section contains the following columns: \textit{Avg}, which shows the average value of the objective function of the problem (the total cost of the STSP) or execution time, \textit{Stan.Dev.}, which represents the standard deviation of the objective function or execution time and, in the sections related to the objective function, an additional column \textit{\%Sol.Inst.} indicates the percentage of instances solved of the 10 instances run. The symbol ``-'' indicates that no solution is found in any of the 10 executions. 

The results demonstrate the clear advantages of the RILP version over SILP, particularly in terms of scalability and computational efficiency for larger instances. Both methods yield identical objective function values across all tested problem sizes, indicating that RILP does not compromise solution quality. However, a significant difference emerges when considering the computational time required for each method. Indeed, RILP consistently outperforms SILP in this regard, providing solutions in considerably less time. A key distinction between the two methods is their ability to handle larger instances. While SILP fails to find solutions for $|V| = 16, 17, 18, 19$, RILP successfully finds solutions for $|V| = 16$ and $|V| = 17$. Notably, for $|V| = 18$, RILP finds a solution 90\% of the time, and for $|V| = 19$, RILP is able to solve the problem once but fails for the next larger instance. This demonstrates RILP's greater scalability, as it consistently manages to solve larger instances where SILP cannot provide any solution.

We then transform both our SILP and RILP formulations into QUBO and evaluate the impact of the PMRA in reducing the problem size. Table~\ref{scalabilità} presents a comparison of the number of variables in the SQUBO and RQUBO formulations, highlighting the effectiveness of the PMRA in reducing the number of variables across different problem sizes.
It consists of the column $|V|$, representing the cardinality of the set $|V|$, followed by \textit{n.var SQUBO} and \textit{n.var RQUBO}, which indicate the number of variables in the SQUBO and RQUBO formulations, respectively. Finally, the \textit{GAP} column shows the percentage reduction achieved.  
Analyzing the results, we observe that the number of variables in the SQUBO formulation grows rapidly as $|V|$ increases. When the PMRA is used, the increase in the number of variables as the size of the set V grows is more contained. In fact, a maximum reduction of 57\% and an average reduction of 48\% are achieved.

\subsection{Results using Simulated Annealing}
We present the outcomes of solving the problem after transforming it into a QUBO formulation. We examine both the SQUBO and RQUBO versions of the STSP, solving them using the SA algorithm. The results obtained from the classical SA solver are shown in Table~\ref{SA}.
This table is divided into four sections: \textit{SA for SQUBO Objective Function}, \textit{SA for SQUBO Time}, \textit{SA for RQUBO Objective Function} and \textit{SA for RQUBO Time}. Each section contains the same columns as the Table~\ref{QUBO}. 

When transforming an ILP problem into a QUBO formulation, the structure of the model changes. As a result, the energy of the QUBO formulation does not directly correspond to the objective function of the original ILP. This discrepancy arises from intrinsic differences in constraints and variable representations between the two approaches. Consequently, the numerical values obtained from the two formulations are not directly comparable.
Our results indicate that RQUBO consistently outperforms SQUBO, yielding solutions that are closer to the optimal compared to those obtained using SA for the SQUBO formulation. In particular, we observe that for both $|V| = 4$ and $|V| = 5$, the solutions obtained with SA for RQUBO are consistently closer to the optimal solution and are found in a shorter time compared to SA for SQUBO. A more pronounced difference emerges for $|V| = 5$, where SA for SQUBO finds a solution in only 30\% of the runs, whereas SA for RQUBO successfully solves 100\% of the instances. Additionally, across all instances where solutions are available, SA for RQUBO consistently requires less computational time than SA for SQUBO, further demonstrating its superior efficiency.

\subsection{Results using Quantum Annealing}
We analyze the results obtained from the QA approach, utilizing both the D-Wave QPU and the D-Wave LeapBQM solver in our experiments. The results presented in Table~\ref{QA with QPU} highlight the performance of the QA approach using the D-Wave QPU for solving both the SQUBO and RQUBO formulations of the STSP problem. The table is organized into four distinct sections: \textit{QA with QPU for SQUBO Objective Function}, \textit{QA with QPU for SQUBO Time}, \textit{QA with QPU for RQUBO Objective Function}, and \textit{QA with QPU for RQUBO Time}. Each of these sections presents data similar to what is shown in Tables~\ref{QUBO} and \ref{SA}, with identical column structures, focusing on the respective objective function and execution time for both the SQUBO and RQUBO formulations.

The RQUBO approach consistently yields better objective function values, bringing the solutions closer to the optimal, minimal value. For $|V| = 4$, the average objective function value for RQUBO is significantly smaller than for SQUBO, demonstrating the RQUBO formulation's ability to converge toward better solutions. 

In terms of execution time, the RQUBO formulation also outperforms SQUBO, with notably faster solution times for smaller instances. However, as the problem size increases (e.g., for $|V| = 5$), the performance of the QPU-based QA approach significantly degrades. This is evident from the much larger average execution time for RQUBO compared to SQUBO, as the system struggles with the increased complexity of the optimization problem. In general, the instances are relatively small, which highlights the increasing complexity of the problem as the size grows.

Table~\ref{QA with hybrid solver} presents the results obtained using the \texttt{LeapBQM} hybrid solver. 
It shows that RQUBO consistently outperforms SQUBO, yielding significantly lower objective function values across all tested problem sizes. For instance, when $|V| = 4$, the average objective function value for SQUBO is 1508,31, while for RQUBO, it is considerably lower at 529,95. This trend persists as the problem size increases: at $|V| = 9$, SQUBO reaches an average value of 57309,63, whereas RQUBO achieves a significantly reduced value of 28061,95. On average, the objective function value for RQUBO is 55\% lower than that of SQUBO, highlighting its superior optimization capability.  

Regarding execution time, RQUBO also demonstrates a slight advantage over SQUBO. While both methods consistently solve all instances, SQUBO generally exhibits slightly faster execution times. Despite this, the substantial reduction in objective function value achieved by RQUBO justifies its slightly higher computational cost, making it the preferred approach for obtaining high-quality solutions.  

\section{Conclusions}\label{sec:conclusions} 
In this work, we have addressed the STSP, a well-known NP-hard problem that is notoriously difficult to solve, by employing a hybrid classic-quantum approach. Specifically, we utilized QA, after transforming the ILP model into a QUBO formulation. We experimentally demonstrated that the number of variables increases significantly as a result of this QUBO transformation. To resolve this problem, we proposed a reduction method called PMRA, which focuses on eliminating unnecessary arcs in the network. We experimentally validated the effectiveness of our method, showing that it consistently improves performance. Subsequently, we solved the problem using both SA and the D-Wave quantum annealer.

We utilized both the D-Wave QPU and the D-Wave \texttt{LeapBQM} solver in our experiments. The \texttt{LeapBQM} consistently outperformed the QPU-based approach, both in terms of execution time and solution quality. While the QPU demonstrated effectiveness for very small problem instances, it was generally less efficient for larger problems. The \texttt{LeapBQM} provided high-quality solutions and showed better scalability, even for larger problem sizes. This makes the \texttt{LeapBQM} a more promising and practical tool for addressing real-world STSP challenges, offering superior performance across a range of problem sizes. Furthermore, in all the tests conducted across all methods, ILP solved with Gurobi, SA and both quantum approaches (QPU and \texttt{LeapBQM}), we observed that the PMRA consistently enhanced the solving performance. This method proved to be effective in improving the computational efficiency and the quality of the solutions, underlining its importance as a significant contribution to solving the STSP.

For future work, a deeper exploration into why QA still provides inferior results compared to the ILP version solved by Gurobi would be highly valuable. Understanding the limitations of quantum approaches in this context could offer important insights and help identify ways to improve their performance. Additionally, further investigation into the potential of QA for solving larger and more complex instances would contribute to the development of more scalable quantum optimization techniques, making them more applicable to real-world problems. Another promising direction for future research involves exploring quadratic formulations. This approach would allow for greater flexibility in modeling and may lead to improved results, especially when handling time-dependent constraints or more intricate network structures. Finally, as the problem scope expands to larger logistics networks and more complex instances, there is potential for further refinements to the PMRA method. This would enable better handling of increasingly complicated scenarios, providing a deeper analysis of the trade-offs between classical and quantum methods.

\section*{Acknowledgments}
Eneko Osaba acknowledges support from the Basque Government through (EXP.2022/01341) (A/20220551) (\textit{Plan complementario de comunicación cúantica}). During the preparation of this work, the authors used Microsoft Copilot to improve the language and readability of the manuscript. After using this tool/service, the authors have reviewed and edited the content as needed and take full responsibility for the content of the publication.

\bibliographystyle{ACM-Reference-Format}
\balance  
\bibliography{bibliography}

\end{document}